\documentclass[linenumbers]{aastex631}
\nolinenumbers

\usepackage{amsmath}
\usepackage[version=3]{mhchem} 

\begin{document}

\title{Resolved convection in hydrogen-rich atmospheres}

\author[0000-0003-0769-292X]{Jacob T. Seeley}
\affiliation{Department of Earth and Planetary Sciences \\
Harvard University \\
Cambridge, MA 02461, USA}

\author[0000-0003-1127-8334]{Robin D. Wordsworth}
\affiliation{Department of Earth and Planetary Sciences \\
Harvard University \\
Cambridge, MA 02461, USA}



\begin{abstract}

In hydrogen-rich atmospheres with low mean molecular weight (MMW), an air parcel containing a higher-molecular-weight condensible can be negatively buoyant even if its temperature is higher than the surrounding environment. This should fundamentally alter the dynamics of moist convection, but the low-MMW regime has previously been explored primarily via one-dimensional theories that cannot capture the complexity of moist turbulence. Here, we use a three-dimensional cloud-resolving model to simulate moist convection in atmospheres with a wide range of background MMW, and confirm that a humidity threshold for buoyancy reversal first derived by \cite{Guillot1995} coincides with an abrupt change in tropospheric structure. Crossing the ``Guillot threshold" in near-surface humidity causes the dry (subcloud) boundary layer to collapse and be replaced by a very cloudy layer with a temperature lapse rate that exceeds the dry adiabatic rate. Simulations with reduced surface moisture availability in the lower atmosphere feature a deeper dry subcloud layer, which allows the superadiabatic cloud layer to remain aloft. Our simulations support a potentially observable systematic trend toward increased cloudiness for atmospheres with near-surface moisture concentrations above the Guillot threshold. This should apply to \ce{H2O} and potentially to other condensible species on hotter worlds. We also find evidence for episodic convective activity and associated variability in cloud cover in some of our low-MMW simulations, which should be investigated further with global-scale simulations.

\end{abstract}



\section{Introduction} \label{sec:intro}

On Earth, convective clouds are positively buoyant because they are warmer than the surrounding air. However, this link between warmth and positive buoyancy is not a universal feature of moist convection in planetary atmospheres. \cite{Guillot1995} recognized that when an atmosphere's condensing gas has a higher molecular weight than the background (non-condensing) gas, positive temperature perturbations can make saturated air \textit{negatively} buoyant. This is not possible on present-day Earth because water vapor is lighter than the background nitrogen-oxygen mixture, which gives rise to the ``virtual effect" that enhances cloud buoyancy \citep[also known as ``vapor buoyancy";][]{Guldberg1876,Yang2020,Seidel2020,Yang2022,Yang2023}. However, the less familiar situation envisioned by \cite{Guillot1995}, in which (anomalously) warm, saturated air is negatively buoyant because it is enriched in the heavier condensing gas, almost certainly arises for clouds made of water, methane, or ammonia in the primarily hydrogen-helium atmospheres of the gas giants in our solar system \citep[e.g.,][]{Leconte2017,Friedson2017}. H$_2$-rich atmospheres with low mean molecular weight may also prevail on smaller terrestrial planets early in their evolution, before they are lost to space \citep{Pierrehumbert2011,wordsworth2012transient,mol2022potential}. Outside our solar system, sub-Neptunes are thought to be one of the most abundant types of planet \citep{Petigura2013,Marcy2014,Winn2015}, and some of these may be water-rich with H$_2$-dominated atmospheres \citep[e.g., ][]{madhusudhan2023carbon,Innes2023}. Despite the abundance of H$_2$-rich atmospheres, however, the potentially unusual dynamics of convective clouds in such atmospheres have not been studied in much detail. How does moist convection operate when positive temperature perturbations do not necessarily produce positive buoyancy and upwelling motion, and what are the implications of such a regime for cloud cover and other observable quantities? 

Here, we use a three-dimensional model that explicitly resolves convection to fully explore the state space between the Earth-like cloud regime and the H$_2$-rich regime. Rather than designing our simulations to mimic the conditions on specific planets \citep[as in, e.g.,][]{Sugiyama2011,Sugiyama2014,Li2019,Leconte2024}, we build a more general understanding of convective dynamics by varying the mean molecular weight of the background gas over a wide range.

In Section~\ref{sec:theory}, we describe the theoretical background to this problem. In Section~\ref{sec:exp}, we describe the experimental setup for our high-resolution cloud-resolving model (CRM) simulations. Results are presented in Section~\ref{sec:results}. In Section~\ref{sec:disc}, we present our conclusions and discuss implications of our work, including possible extension to other condensing species and to exoplanet observations.

\section{Theoretical Background}\label{sec:theory}

The thermodynamic properties of dry air in our model are solely dictated by its specific heat at constant volume, $c_a$ (J~Kg$^{-1}$~K$^{-1}$), and its specific gas constant, $R_a$ (J~kg$^{-1}$~K$^{-1}$); the specific heat at constant pressure of dry air is then $C_a=c_a+R_a$, as usual. In general, the specific gas constant for substance $x$ is related to the universal gas constant $R=8.3145$ (J~mol$^{-1}$~K$^{-1}$) by $R_x = R/\mu_x$, where $\mu_x$ (kg/mol) is the mean molar mass of substance $x$. From the equipartition theorem, the specific heat at constant volume for a diatomic molecule is $c_x = \frac{5}{2}R_x=5 R/(2\mu_x)$ and the specific heat at constant pressure is then $C_x = 7 R/(2 \mu_x)$. In our experiments, we vary the mean molar mass of the dry-air component, $\mu_a$, from that of N$_2$ ($\simeq28$ g/mol) to that of H$_2$ ($\simeq2$ g/mol), and account for this in the numerical model by varying the specific heat and gas constant for the dry-air component according to the above basic thermodynamic relations for an ideal diatomic gas. Experiments with mean molar masses between N$_2$ and H$_2$ may be considered as weighted mixtures of the two end-member gases, or as primordial \ce{H2}-He atmospheres with metallicity increased by some factor. For example, assuming an atmosphere with 100$\times$ solar metallicity results in a mean molecular weight of approximately 4 g/mol  for the exoplanet GJ 1214b \citep{drummond2018effect}.

\begin{figure}[ht!]
\centerline{\includegraphics[width=4in]{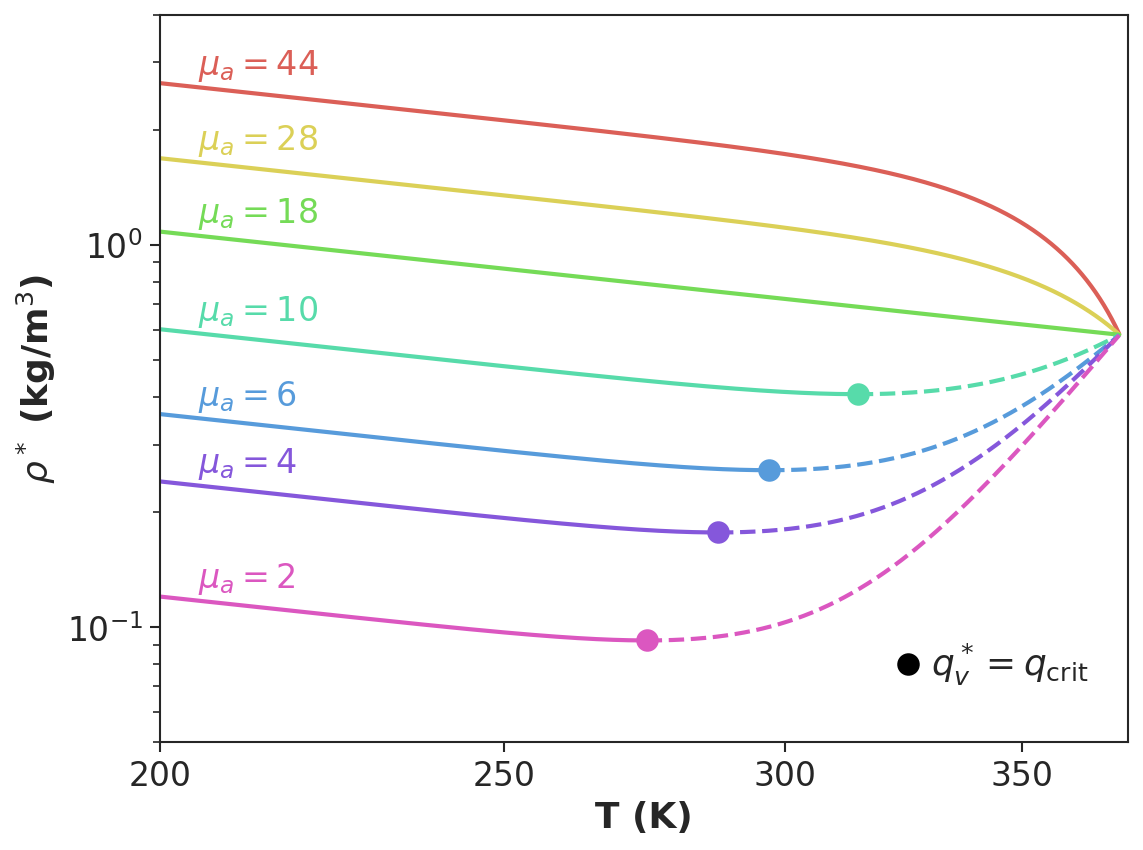}}
\caption{The saturation density $\rho^*=p \mu^*/(R T)$ for a mixture of water vapor and a background gas with molar mass $\mu_a$ varying from 44 g/mol to 2 g/mol. The total pressure is held fixed at 10$^5$ Pa, and the temperature ranges from 200 K to the boiling point of water at this pressure. For cases in which the background gas is lighter than water vapor, the portions of the curves for which $\rho^*$ increases with temperature are plotted with dashed lines, and the point where the saturation specific humidity $q_v^*$ equals the critical humidity threshold $q_\mathrm{crit}$} (eqn. \ref{eq:qcrit}) is marked with a solid circle.
\label{fig:rhostar}
\end{figure}

As a theoretical preliminary, we now derive the conditions under which anomalous warmth makes saturated air negatively buoyant \citep{Guillot1995}. Consider a parcel of air at temperature $T$ and pressure $p$; the total pressure is $p=p_a + p_v$, where $p_a$ is the partial pressure of dry air and $p_v$ is the vapor pressure of the condensing component. The ideal gas law reads
\begin{equation}
    \rho = \frac{p}{R_m T},
\end{equation}
\noindent where $R_m = q_v R_v + (1-q_v)R_a$ is the specific gas constant for two-component moist air, the (condensing) vapor mass fraction is $q_v$, the dry air mass fraction is $q_a=1-q_v$, and $R_v$ is the specific gas constant of vapor. Expressing the moist-air gas constant in terms of the mean molar mass, we can rewrite the ideal gas law as
\begin{equation}
    \rho = \frac{p \mu}{R T},\label{eq:ideal2} 
\end{equation}
\noindent where the mean molar mass $\mu$ is given by
\begin{equation}
    \mu = \frac{\mu_a}{1 - \alpha q_v}, \label{eq:mu}
\end{equation}
\noindent and where we have defined the fractional molar mass difference
\begin{equation}
    \alpha \equiv \frac{\mu_v - \mu_a}{\mu_v}.
\end{equation}
Denoting quantities at saturation with a superscript asterisk, we can define a ``saturation density", $\rho^*=p \mu^*/(R T)$, which is the density of an air parcel at temperature $T$ and pressure $p$ when the condensible component is at saturation (i.e., $p_v = p_v^*$ and $q_v = q_v^*$). In Figure \ref{fig:rhostar}, we plot $\rho^*$ for a parcel of air at 1 bar (10$^5$ Pa) pressure over a wide range of temperature, and for varying background molar mass. The condensible in this case is water vapor, the thermodynamics of which we model with a set of approximations \citep{Romps2021} that are standard in Earth meteorology\footnote{The approximations are that the vapor behaves as an ideal gas, the heat capacities of all phases are constant, and the condensates have zero specific volume.}, neglecting the solid phase for simplicity (this does not affect our results). We see that for $\alpha\leq0$ (the cases with $\mu_a=44$, 28, or 18), increasing temperature always decreases saturation density. Physically, if the molar mass of vapor is less than that of dry air, then as temperature increases the saturated air parcel is increasingly dominated by the lighter vapor molecules, and its mean molecular weight will decrease. Mathematically, $\mu^*=\mu_a/(1 - \alpha q_v^*)$ is a monotonically decreasing function of temperature when $\alpha < 0$, because the denominator of equation (\ref{eq:mu}) increases with the saturation specific humidity $q_v^*$.  This means that in these cases with $\alpha\leq0$, an anomalously warm, saturated parcel will always be less dense than surrounding air at the same pressure, and therefore the warm parcel will be buoyant\footnote{This conclusion holds even if the surrounding air is not saturated (as is likely for clear air), because the buoyancy of the saturated parcel would be further enhanced by the lower vapor content of the environment.}.

For $\alpha > 0$ (the cases in Fig. \ref{fig:rhostar} with $\mu_a=10$, 6, 4, and 2), the situation is more interesting. In this case $\mu^*$ is an increasing function of $T$, and the growth of $\mu^*$ can (in principle) overpower the effect of increasing $T$ on density. At cool temperatures, when $q_v^*$ is small, the direct effect of $T$ on density wins out, and $\rho^*$ decreases with $T$ as in the $\alpha\leq0$ case. But above a critical threshold --- marked by the solid circles in Figure \ref{fig:rhostar} --- the molar mass effect wins out, and $\rho^*$ becomes an increasing function of $T$ (dashed line segments in Fig. \ref{fig:rhostar}). This is the unfamiliar regime in which anomalous warmth makes a saturated air parcel more dense, and therefore negatively buoyant\footnote{Again, if the environment is not saturated, this only enhances the negative buoyancy of the anomalously warm parcel.}.

The critical threshold above which $\rho^*$ increases with $T$ can be derived by taking the logarithm of the ideal gas law (eqn. \ref{eq:ideal2}) and differentiating at fixed pressure:
\begin{equation}
    \frac{d \ln \rho^*}{d \ln T}\Bigr|_{p} = \frac{d \ln \mu^*}{d \ln T}\Bigr|_p - 1.
\end{equation}
\noindent Hence the condition
\begin{equation}
    \frac{d \ln \mu^*}{d \ln T}\Bigr|_p = 1
\end{equation}
\noindent identifies the point at which the derivative of $\rho^*$ with respect to $T$ changes sign. By the chain rule,
\begin{equation}
    \frac{d \ln \mu^*}{d \ln T}\Bigr|_p = \left(\frac{d\ln \mu^*}{d \ln q_v^*}\right) \left(\frac{\partial \ln q_v^*}{\partial \ln T}\right),
\end{equation}
\noindent where the saturation specific humidity is given as a function of temperature and pressure by
\begin{equation}
    q_v^* = \frac{\epsilon p_v^*}{p + (\epsilon-1)p_v^*},
\end{equation}
\noindent and where $\epsilon=\mu_v/\mu_a$. From the definition of $\mu$ (eqn. \ref{eq:mu}), we obtain
\begin{equation}
    \frac{d \ln \mu^*}{d \ln q_v^*} = \frac{\alpha q_v^*}{1 - \alpha q_v^*}.
\end{equation}
\noindent Furthermore, from the definition of $q_v^*$, we obtain
\begin{equation}
    \frac{\partial \ln q_v^*}{\partial \ln T} = \left(1 - \alpha q_v^*\right)\frac{d \ln p_v^*}{d \ln T}.
\end{equation}
\noindent Differentiating the standard Rankine-Kirchoff expression for saturation vapor pressure \citep{Romps2021} yields
\begin{equation}
\frac{d \ln p_v^*}{d \ln T} = \frac{L(T) \mu_v}{R T},
\end{equation}
\noindent where $L(T)=E_{0v} + R_v T + (c_{vv}-c_{vl})(T-T_0)$ is the temperature-dependent latent heat of condensation; in this expression, $E_{0v}$ is the internal energy difference between vapor and liquid at the reference temperature $T_0$ (often taken to be the triple point when the condensible is water), while $c_{vv}$ and $c_{vl}$ are the specific heat at constant volume of the condensible in vapor and liquid form, respectively. Putting it all together, we find
\begin{equation}
    \frac{d \ln \mu^*}{d \ln T}\Bigr|_p = \alpha q_v^* \frac{L(T) \mu_v}{R T}.
\end{equation}
\noindent Setting the above equal to 1 and solving for $q_v^*$, we recover a result from \cite{Guillot1995}:
\begin{equation}
    q_\mathrm{crit} = \frac{R T}{(\mu_v - \mu_a) L(T)}. \label{eq:qcrit}
\end{equation}
\noindent Note that equation (\ref{eq:qcrit}) only applies when $\mu_v>\mu_a$. The solid circles in Figure \ref{fig:rhostar} mark the condition $q_v^*=q_\mathrm{crit}$, confirming that equation \ref{eq:qcrit} predicts the transition to the ``Guillot regime" in which anomalous warmth corresponds to negative buoyancy for saturated air. 

\section{Experimental methods}\label{sec:exp}


\subsection{Cloud-resolving model}
We simulated nonrotating radiative-convective equilibrium on doubly periodic domains with the cloud-resolving model DAM \citep{Romps2008}, with an experimental setup that is overall similar to that of \cite{Seeley2023}. For our core experiments (see Section \ref{sec:exps}), the vertical grid had a variable spacing that transitions smoothly from $\Delta z=50$ m below an altitude of 550 m to $\Delta z=1000$ m at altitudes between 10 km and the model top, which was placed at a variable height because our simulated atmospheres vary widely in atmospheric scale height and geometric depth. The core experiments all used a horizontal resolution of $\Delta x =\Delta y= 2$ km and a horizontal domain size of $L_x=L_y=96$ km. Some aspects of simulated convection, such as cloud fraction and precipitation efficiency, are sensitive to vertical and horizontal resolution \citep[e.g.,][]{Jeevanjee2022,Jenney2023}, but we do not explore this sensitivity in detail in this work. However, we did run one simulation with higher resolution ($\Delta x =\Delta y= 250$ m) on a quasi-2D domain ($L_x=750$ m, $L_y=128$ km); for this higher-resolution simulation, the vertical grid spacing transitioned smoothly from $\Delta z=25$ below 650 m altitude, to $\Delta z=100$ m at 1400 m altitude, and then to $\Delta z=500$ m at and above 6570 m altitude. The vertical dimensions and number of vertical levels for all of the simulations are summarized in Table \ref{tab:zdims}.

The model time step was $\Delta t=20$~s, sub-stepped to satisfy a Courant-Friedrichs-Lewy (CFL) condition. Since DAM has a fully-compressible dynamical core, it includes acoustic modes and fast gravity waves. The speed of sound in air is 
\begin{equation}
    v_\text{sound} = \sqrt{\frac{\gamma RT}{\mu}}, \label{eq:v_snd}
\end{equation}
\noindent where $\gamma$ is the adiabatic index (equal to 7/5 for a diatomic gas). In DAM, the number of sub-steps that are required to propagate these fast modes during one large model time step is determined by a parameter $v_\text{sound,DAM}$ that corresponds to a typical speed of sound in the simulated atmosphere. We account for the effect of varying molar mass $\mu$ by adjusting the parameter $v_\text{sound,DAM}$ in our simulations according to equation \ref{eq:v_snd}. Since $\mu$ and $T$ vary over the depth of the model atmosphere, we set $v_\text{sound,DAM}$ to the average of the value of $v_\text{sound}$ at the bottom and top of the model when the simulation is initialized. 

Consistent with our focus on convective dynamics, we used a simplified treatment of radiative transfer in our simulations \citep[as in, e.g.,][]{Tan2021,Seeley2023}. We prescribed an idealized column-integrated tropospheric radiative cooling of 200 W/m$^2$, with the cooling rate (in K/day) distributed uniformly in the vertical between the surface and the level in the atmosphere where the domain-average temperature reached a prescribed ``tropopause" temperature of 200 K. At altitudes above this diagnosed tropopause, temperatures were simply nudged to the tropopause temperature on a timescale of 6 hours. This idealized radiative forcing, while allowing us to focus on the effect of $\mu_a$ on convective dynamics, does have its limitations. Fixing the bulk tropospheric radiative cooling constrains the mean enthalpy flux from the surface and the convective mass flux, but in reality these quantities may vary as a function of $\mu_a$. Follow-up work with a fully-interactive radiative transfer scheme is warranted.

All of our simulations used the simplified microphysics scheme described in previous work \citep{Seeley2023}; briefly, the simplified microphysics scheme tracks three bulk classes of condensible substance: vapor, non-precipitating cloud liquid, and rain. The saturation adjustment routine prevents relative humidity from exceeding 100\% and evaporates cloud condensate in subsaturated air, while conversion of non-precipitating cloud condensate to rain is modeled as autoconversion on a fixed timescale; evaporation of rain in subsaturated air is also given a prescribed timescale, and rain is given a fixed freefall speed. All microphysical parameters are identical to those used in \cite{Seeley2023}. 

We assumed a fixed sea surface temperature (SST) in all simulations, and surface fluxes of sensible heat and moisture were modeled with the same standard bulk aerodynamic formulae used in \cite{Seeley2023}. Specifically, the surface latent and sensible heat fluxes (LHF and SHF) were given by

\begin{equation}
    \mathrm{LHF}(x,y) = \rho_1(x,y) C_D\sqrt{u_1(x,y)^2 + v_1(x,y)^2 + V^2} L_e  \left[ \beta q_s^* - q_1(x,y)\right]; \label{eq:LHF}
\end{equation}

\begin{equation}
    \mathrm{SHF}(x,y) = \rho_1(x,y) C_D\sqrt{u_1(x,y)^2 + v_1(x,y)^2 + V^2} c_p  \left[ \mathrm{SST} - T_1(x,y)\right], \label{eq:SHF}
\end{equation}
\noindent where $\rho_1$, $q_1$, $u_1$, $v_1$, and $T_1$ are the density, specific humidity, horizontal winds, and temperature at the first model level, $C_D=1.5\times10^{-3}$ is a drag coefficient, $V=5$ m/s is a background ``gustiness'', $L_e$ is the latent heat of evaporation, $c_p$ is the specific heat capacity at constant pressure of moist air, $\beta$ is a surface wetness parameter ranging from 0 to 1, and $q_{s}^*$ is the saturation specific humidity at the sea surface temperature and surface pressure. The time-mean surface enthalpy flux is constrained by the (imposed) tropospheric radiative cooling, so the values of $C_D$ and $V$ determine the near-surface air-sea enthalpy disequilibrium but do not otherwise affect our results. While the parameterization of surface fluxes given by equations \ref{eq:LHF}-\ref{eq:SHF} is commonly employed for oceanic convection in Earthlike atmospheres, for hydrogen-dominated atmospheres it is more typical to assume a semi-infinite atmosphere with a fixed mixing ratio of the condensible species at the bottom.\footnote{The recent study of \cite{Markham2022} has developed the theory of hydrogen atmospheres overlying a condensible ocean.} We note that equation \ref{eq:LHF} effectively relaxes the near-surface humidity to the value $\beta q_s^*$ on a timescale set by the drag coefficient and near-surface winds; therefore, in the limit of vanishing relaxation timescale, our formulation of the lower boundary condition should converge to the fixed-mixing-ratio assumption. We do not expect our choice of surface flux parameterization with non-zero relaxation timescale to bias our results in hydrogen-dominated atmospheres.

In one of our experiment suites, we vary the surface wetness by changing the parameter $\beta$ in the bulk aerodynamic formula for latent heat flux, as in \cite{Cronin2019}. Reducing $\beta$ from the value of 1 (appropriate for pure water) limits the availability of moisture at the surface.


\begin{figure}[ht!]
\centerline{\includegraphics[width=\textwidth]{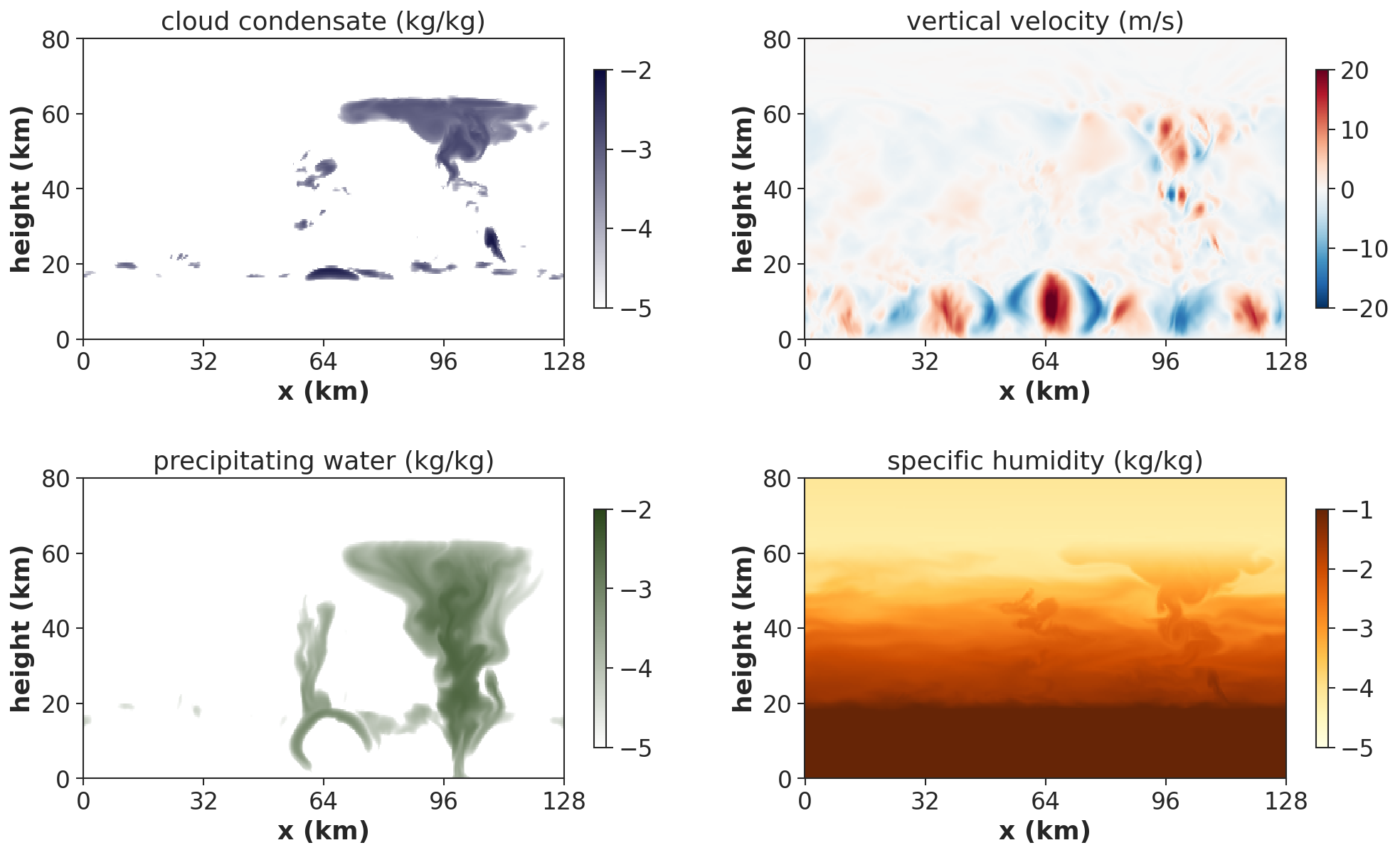}}
\caption{A snapshot of the \textit{DeepBL\_hr} simulation, showing (from upper left, going clockwise): cloud condensate, vertical velocity, specific humidity, and precipitating water. The colorbar for all fields except for vertical velocity is the base-10 logarithm of the water mass fraction (i.e., a value of -5 means 10$^{-5}$ kg/kg.) Note that this simulation was performed on a quasi-2D grid, and with a higher resolution than the rest of our simulations. An animation of this simulation is available in the Supplemental Movie.}
\label{fig:deepBL_hr}
\end{figure}

\subsection{Experiments}\label{sec:exps}

We conducted five core suites of experiments. In the first (\textit{VaryMu}), we simulated atmospheres with a surface pressure of 10$^5$ Pa over a surface temperature of 300 K, with background gas molar mass $\mu_a$ varying from that of N$_2$ (28 g/mol) to that of pure H$_2$ (2 g/mol); note that this range includes the solar composition value \citep[$\simeq$2.5 g/mol;][]{Asplund2009}. The case with $\mu_a=28$ g/mol is quite similar to the current conditions in Earth's tropics, with lower values of $\mu_a$ representing an otherwise Earthlike configuration but with an increasing proportion of the background gas being the lighter molecule H$_2$. We also repeated this experimental setup for surface temperatures of 280 K and 320 K (\textit{VaryMu\_280} and \textit{VaryMu\_320}, respectively). Our fourth core experiment suite (\textit{DeepBL}) used a surface temperature of 350 K and a surface wetness parameter of $\beta=0.2$, with background gas molar mass varying from $\mu_a=4$ g/mol to 14 g/mol. Our final core experiment suite (\textit{VaryBeta}) was branched from the \textit{DeepBL} case with $\mu_a=4$ g/mol, with surface wetness $\beta$ increased from 0.2 to 0.4 and 0.6. In addition to these core experiment suites, we repeated the \textit{DeepBL} case with $\mu_a=6$ g/mol using the higher model resolution on the quasi-2D domain described above (\textit{DeepBL\_hr}). To give a sense of the complexity of the turbulent convective dynamics resolved by our simulations, a snapshot of the \textit{DeepBL\_hr} simulation is shown in Figure \ref{fig:deepBL_hr}; an animation of this simulation is also available in the Supplemental Movie.

\section{Results}\label{sec:results}
\begin{figure}[ht!]
\centerline{\includegraphics[width=6in]{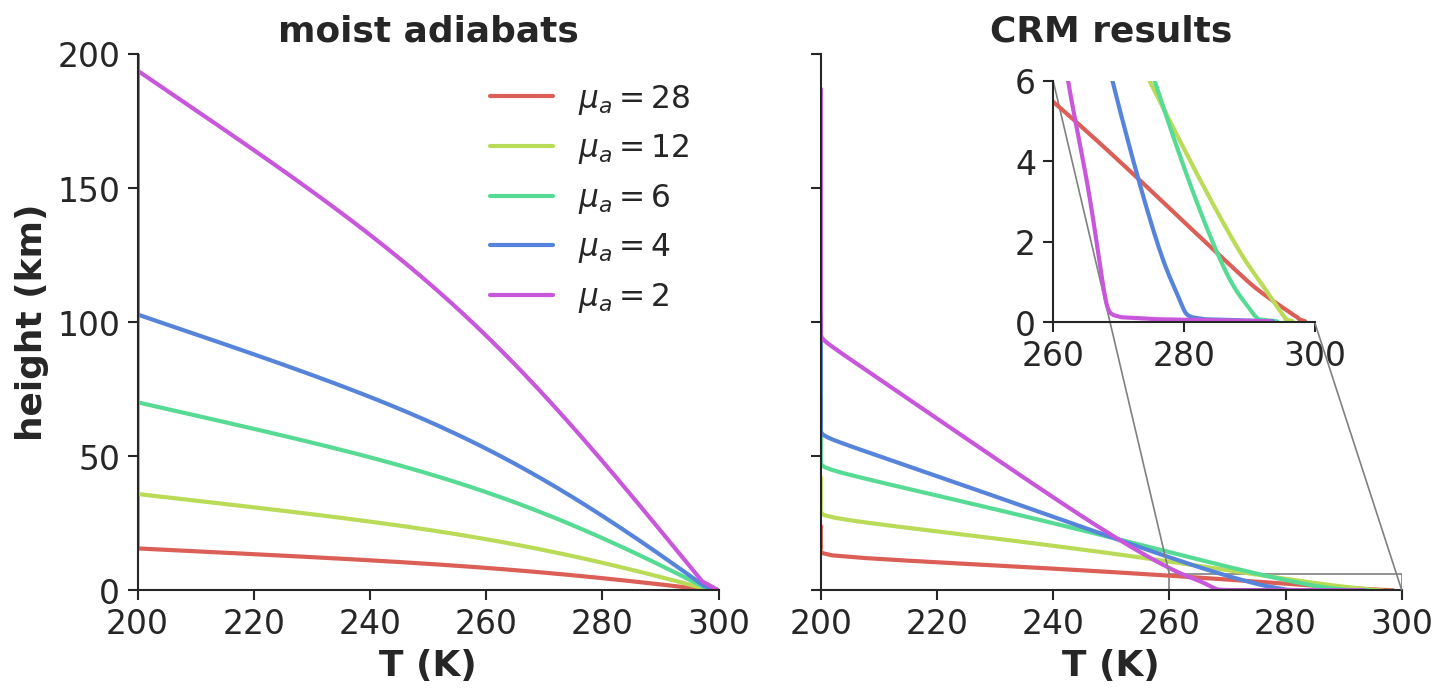}}
\caption{Temperature profiles for atmospheres with a surface temperature of 300 K, surface pressure of 10$^5$ Pa, and with varying background gas molar mass $\mu_a$. The left panel shows idealized moist adiabats, while the right panel shows the steady-state temperature profiles from our cloud-resolving simulations. The inset focuses on the lowermost 6 km of the CRM simulations.}
\label{fig:Tprofs}
\end{figure}

We first present results from the \textit{VaryMu} experiment. In Figure \ref{fig:Tprofs}, we plot areal-mean temperature profiles averaged over the last 50 days of the simulations for atmospheres with a surface temperature of 300 K, surface pressure of 10$^5$ Pa, and with varying background gas molar mass $\mu_a$; For comparison with our model results, the left panel of this figure shows the idealized moist adiabats that are often used to represent tropospheric temperature structure in simplified 1-dimensional climate models \citep[e.g.,][]{Koll2019}. For the adiabats, the dominant effect of varying $\mu_a$ is the increase in the atmospheric scale height $R T/(\mu g)$, where $g$ (m/s$^2$) is the planetary gravitational constant. Clearly, the temperature lapse rate (as measured in K/km) is much reduced in atmospheres with a low mean molecular weight. As emphasized by \cite{Koll2019}, this may allow hydrogen-rich atmospheres to store much more water vapor at a given surface temperature than atmospheres with a larger $\mu$ and smaller scale height.

Our CRM simulations (Fig. \ref{fig:Tprofs}, right panel) tell a more complicated story. For the cases with $\mu_a=28$ and 12 g/mol, the moist adiabats are a reasonable approximation to the steady-state temperature profile of the simulations. But for the more H$_2$-rich cases (i.e., those with $\mu_a\leq6$), the moist adiabats dramatically overpredict temperatures throughout the troposphere. As can be seen in the inset figure, this is because these H$_2$-rich cases contain highly superadiabatic layers near the surface, in which temperature falls by a few tens of kelvin within a few hundred meters. (In the $\mu_a=2$ case, the lapse rate in the lowermost 250 m is approximately 120 K/km.) These rapid declines in temperature are, by the Clausius-Clapeyron relation, associated with rapid declines in water vapor content. As a result, when the temperature structure in these H$_2$-rich cases reverts to an approximately moist adiabatic profile above the superadiabatic layers, these moist adiabats are rooted in a much lower ``effective'' surface temperature. For example, for the $\mu_a=2$ case, it is as if the surface temperature is nearly 270 K, rather than 300 K.

\begin{figure}[hb!]
\centerline{\includegraphics[width=4in]{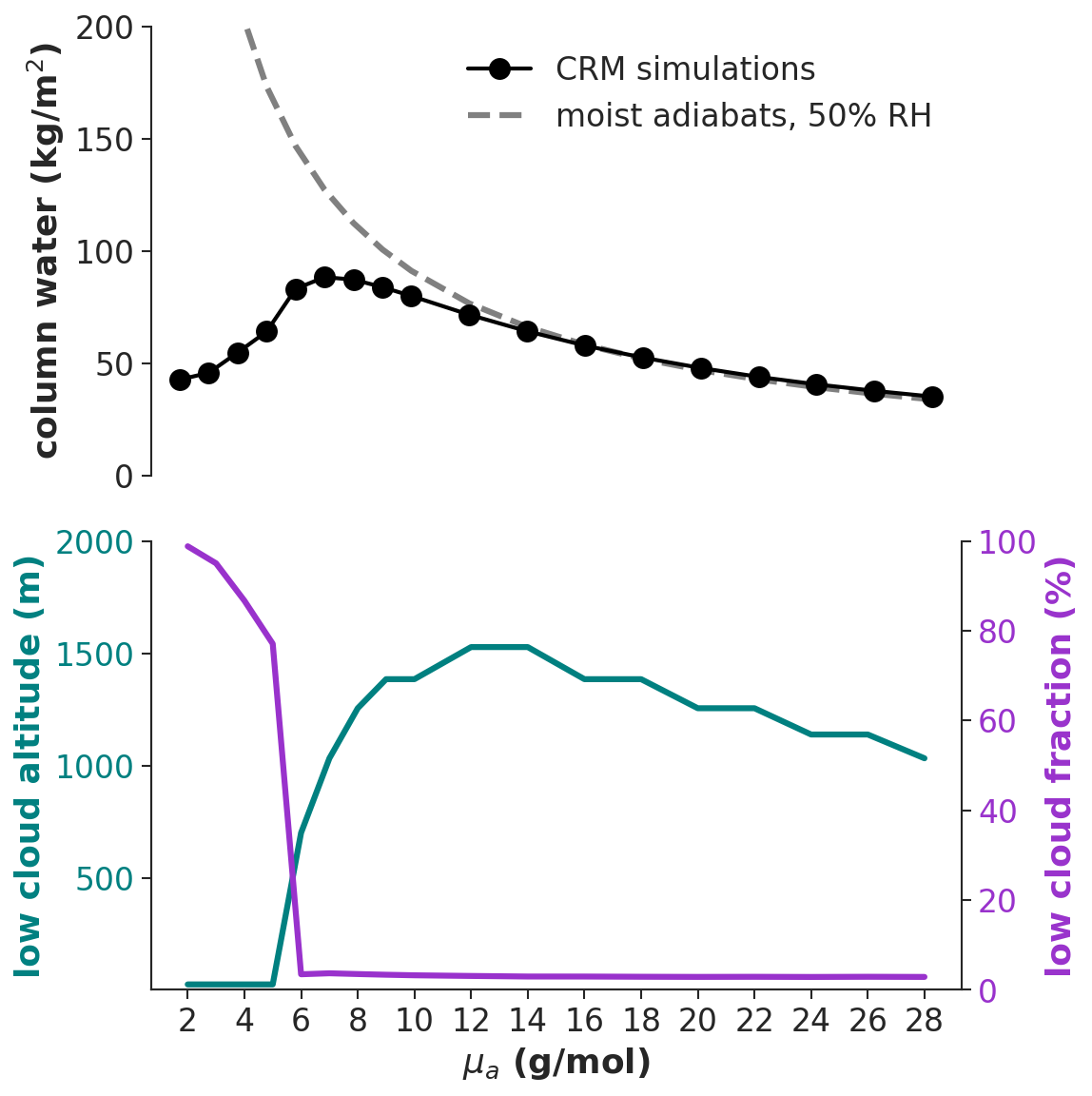}}
\caption{(Top panel) Column-integrated water vapor in the CRM simulations from the \textit{VaryMu} experiment, in comparison to the same quantity calculated from idealized moist-adiabatic atmospheres with 50\% relative humidity. (Bottom panel) Low ($z\leq5$ km) cloud altitude (i.e., the altitude of the peak in cloud fraction below 5 km) and amount (i.e., the magnitude of the low cloud peak) in the \textit{VaryMu} experiment.}
\label{fig:CWV_cld}
\end{figure}

The formation of these superadiabatic near-surface layers, and the resulting dramatic decrease in tropospheric temperatures, corresponds to a notable decrease in the amount of water vapor stored in the atmosphere (Fig. \ref{fig:CWV_cld}). Unlike idealized moist-adiabatic atmospheres, which monotonically increase in water-vapor content as $\mu_a$ is decreased, the water vapor content of the CRM simulations at first increases, but then begins to decrease for $\mu_a\leq7$ g/mol. Associated with the formation of the superadiabatic near-surface layers is a dramatic restructuring of the low cloud cover in our simulations: the peak in low cloud fraction jumps from a few percent for $\mu_a>6$ g/mol to nearly 100\% for $\mu_a=2$ g/mol, while the altitude of the low cloud peak drops from above 1.5 km to the near-surface model level at 25 m altitude. Essentially, the dry-adiabatic sub-cloud layer collapses entirely in our simulations with $\mu_a\leq5$ g/mol, and is replaced by a shallow, very cloudy layer in which temperature and moisture content drop off superadiabatically.

\begin{figure}[ht!]
\centerline{\includegraphics[width=4.5in]{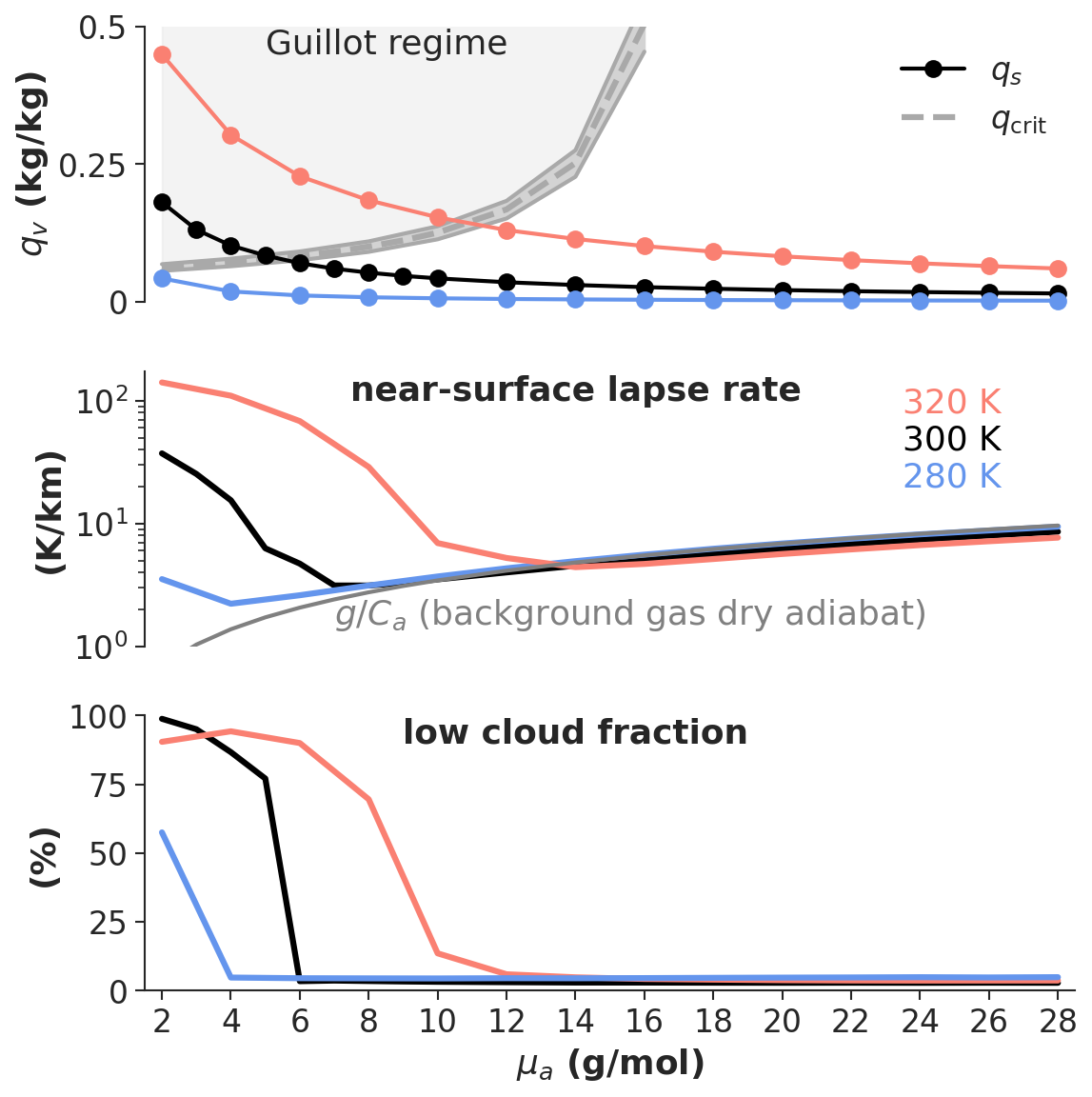}}
\caption{Several quantities from the \textit{VaryMu} experiments at surface temperatures of 280 K (blue), 300 K (black), and 320 K (red). (Top panel) Near-surface specific humidity $q_s$. The critical humidity threshold $q_\mathrm{crit}$ is also plotted with a gray dashed line for the 300 K case; the weak direct temperature dependence of $q_\mathrm{crit}$ (eqn. \ref{eq:qcrit}) is indicated by the two thin gray lines that plot this same quantity for 280 K and 300 K. (Middle panel) Near-surface (50 m $\leq z \leq 250$ m) lapse rate, with the dry-adiabatic lapse rate for background gas molar mass $\mu_a$ also shown as a reference. (Bottom panel) Magnitude of the low cloud fraction peak.}
\label{fig:guillot}
\end{figure}

Why does the dry boundary layer get replaced by a cloudy superadiabatic layer for $\mu_a\leq5$ g/mol? In the top panel of Figure \ref{fig:guillot}, we plot the near-surface specific humidity $q_s$ (kg/kg) for the \textit{VaryMu}, \textit{VaryMu\_280}, and \textit{VaryMu\_320} experiments. We also plot the critical humidity threshold, $q_\mathrm{crit}$, that marks the boundary of the Guillot regime. As $\mu_a$ is reduced, two things happen: 1) $q_s$ in our simulations increases, because the near-surface vapor density is largely controlled by the (invariant) surface temperature while the dry-air density decreases along with the decrease in $\mu_a$; and 2) $q_\mathrm{crit}$ decreases, because less vapor is required to cross into the Guillot regime for lighter background molar masses (eqn. \ref{eq:qcrit}). 

Figure \ref{fig:guillot} clearly indicates that crossing into the Guillot regime (i.e., $q_s\geq q_\mathrm{crit}$) is associated with a dramatic shift in near-surface atmospheric structure: the near-surface (50 m $\leq z \leq 250$ m) lapse rate deviates from the background gas' dry-adiabatic lapse rate of $g/C_a$ and becomes highly superadiabatic, while the magnitude of the low-level cloud fraction increases from a few percent to nearly 100\%. Since the \textit{VaryMu\_320} experiment is the warmest, it has the highest near-surface specific humidity at a given value of $\mu_a$, and therefore crosses into the Guillot regime  at a higher value of $\mu_a$ than the cooler experiments. Indeed, the \textit{VaryMu\_280} experiment only comes close to the $q_s = q_{crit}$ boundary for the lightest background molar mass, $\mu_a=2$ g/mol.

To understand the unusual near-surface atmospheric structure in the Guillot regime, it is useful to imagine an atmosphere that initially has the standard dry-adiabatic subcloud layer underlying a moist troposphere. If this atmosphere has $q_s>q_\mathrm{crit}$, dry convection in the boundary layer will carry air with vapor content $q_s$ up to the lifting condensation level (cloud base), as usual, but condensation of vapor will produce anomalously warm, \textit{negatively buoyant} cloudy parcels. Such parcels will tend to sink back down rather than turn into moist updrafts that can ascend through the troposphere, and so the moist troposphere above the lifting condensation level will continue to undergo radiative cooling. This continued radiative cooling only further enhances the negative buoyancy experienced by cloudy parcels at the lifting condensation level. Hence a larger and larger drop in temperature (and, by Clausius-Clapeyron, drop in moisture content) develops at the top of dry subcloud layer. As this sharp (superadiabatic) drop in temperature develops --- and the associated gradient in $q_v$, set by saturation, also sharpens ---  it is mixed down to the surface by turbulence and numerical diffusion, such that eventually no dry sub-cloud layer remains at all. This process continues until the temperature and moisture have fallen enough at the top of the superadiabatic layer that the atmosphere is no longer in the Guillot regime, at which point parcels that undergo condensational heating can once again turn into moist updrafts.

In the \textit{VaryMu} experiments, the dry subcloud layer disappears entirely once the atmospheres cross into the Guillot regime. But might this be a result of how shallow --- roughly $\mathcal{O}(1$ km) in depth --- the dry subcloud layer is \textit{before} the atmosphere crosses into the Guillot regime? If the boundary layer were deeper to begin with, might a superadiabatic layer develop aloft and remain distinct from the dry subcloud layer, as postulated in \cite{Innes2023} and \cite{Leconte2024}? To test this idea, we conducted an additional experiment suite (named \textit{DeepBL}) with reduced moisture availability at the surface, which has the effect of reducing the relative humidity of near-surface air and raising the lifting condensation level to higher altitudes \citep[e.g.,][]{Cronin2019,Fan2021,McKinney2024}. Specifically, we set the parameter $\beta$ to 0.2 in equation \ref{eq:LHF}, while also increasing the surface temperature to 350~K so that even with lower near-surface relative humidity our simulation is humid enough to be in the Guillot regime. These \textit{DeepBL} simulations with reduced deep moisture are closer to how convection is expected to behave on sub-Neptune exoplanets that do not possess a phase transition to a liquid ocean, or where species other than \ce{H2O} are condensing
\citep[e.g.,][]{Madhusudhan2021,Misener2022,Innes2023}.

\begin{figure}[ht!]
\centerline{\includegraphics[width=6in]{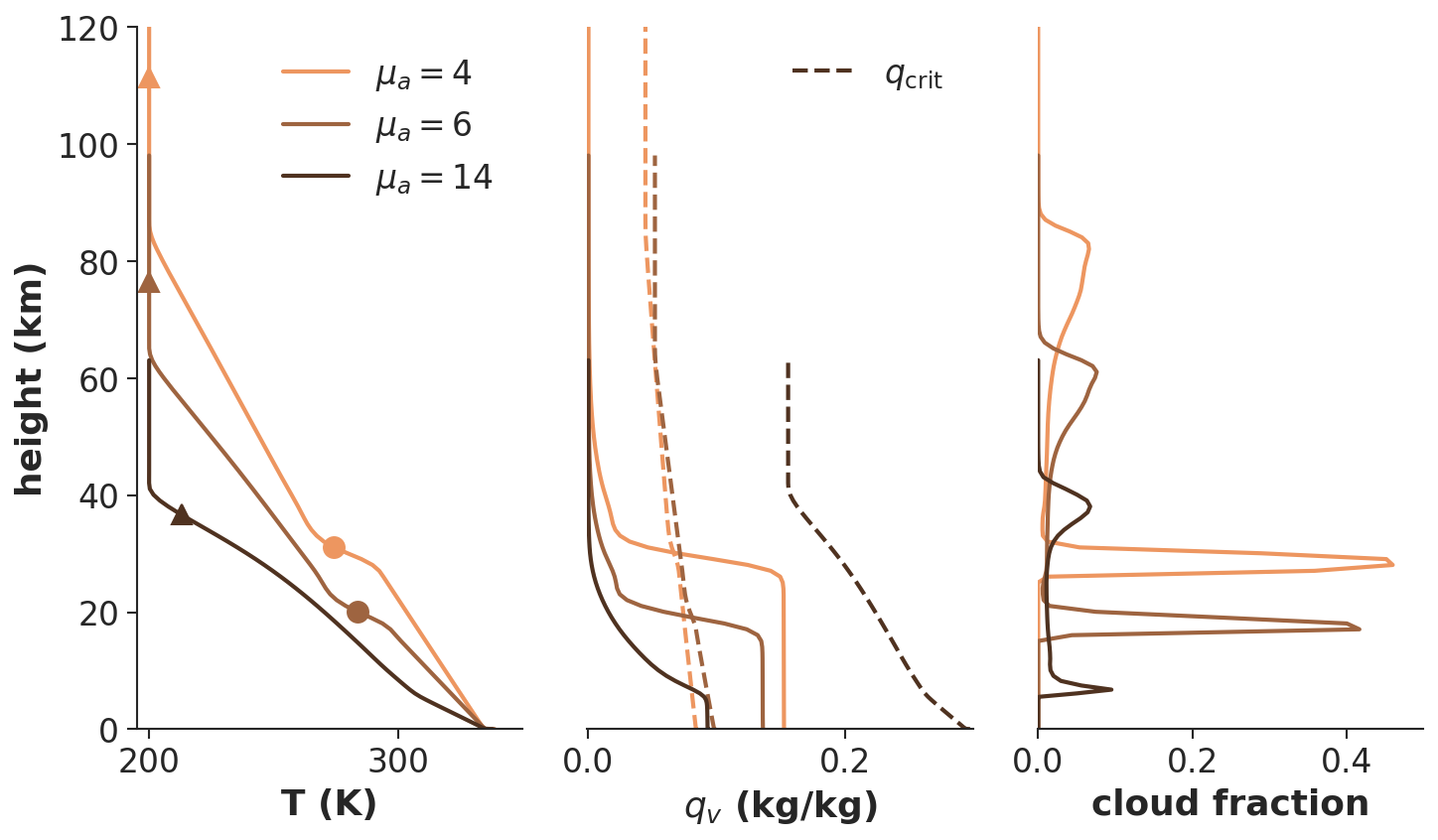}}
\caption{Results from the \textit{DeepBL} experiment for $\mu_a=$ 4, 6, and 14 g/mol. (Left) Temperature profiles, with the altitude at which $q_v=q_\mathrm{crit}$ marked with a solid circle. The $p=0.1$ bar altitude for each simulation is marked with a triangle. (Middle) Specific humidity profiles; the profiles of $q_\mathrm{crit}$ are indicated by the dashed lines). (Right) Cloud fraction profiles.}
\label{fig:deepBL}
\end{figure}

The results shown in Figure \ref{fig:deepBL} confirm that a superadiabatic layer can indeed form aloft and remain distinct from the dry subcloud layer. In the \textit{DeepBL} experiment with $\mu_a=4$ g/mol, the dry subcloud layer is approximately 30 km deep, owing to both the low relative humidity of the near-surface air and the relatively small dry lapse rate of an atmosphere with a light background molar mass. In this simulation, the atmosphere is in the Guillot regime at the lifting condensation level, with $q_v$ being several times larger than $q_\mathrm{crit}$. As a result, the atmosphere does not transition to a moist adiabatic thermal stratification at the lifting condensation level, as in the classic runaway greenhouse temperature profiles \citep[e.g., Fig. 1 of][]{Kasting1988}, but instead transitions to a highly superadiabatic layer in which specific humidity falls by about an order of magnitude \citep[a qualitatively similar, but larger, drop in humidity was theorized by][]{Innes2023}. This layer is very cloudy, with a peak cloud fraction of about 50\%. Only after the humidity has fallen below $q_\mathrm{crit}$ does the superadiabatic layer terminate, transitioning to a second shallow dry-adiabatic layer with approximately uniform $q_v$ and no cloud fraction \citep[as also found in][]{Leconte2024}. Above this secondary dry-adiabatic layer, there is a moist tropospheric layer with cloud fraction of $\mathcal{O}(1-10)$\%. Although there is condensation in this moist layer, the lapse rate is close to the dry-adiabatic lapse rate because the specific humidity above the superadiabatic layer is quite low. A qualitatively similar tropospheric structure emerges in the \textit{DeepBL} case with $\mu_a=6$ g/mol, which is also in the Guillot regime. However, the case with $\mu_a=14$ g/mol is not sufficiently humid to be in the Guillot regime (Fig. \ref{fig:deepBL}, center panel), and this simulation has the more Earth-like two-layer tropospheric structure, with a dry subcloud layer overlain by a moist layer. Note that, in these simulations, the temperature structure and cloud cover of the troposphere mainly lie below the $p=0.1$ bar level, which may make it hard to observe these features in transit spectroscopy.

\begin{figure}[ht!]
\centerline{\includegraphics[width=5in]{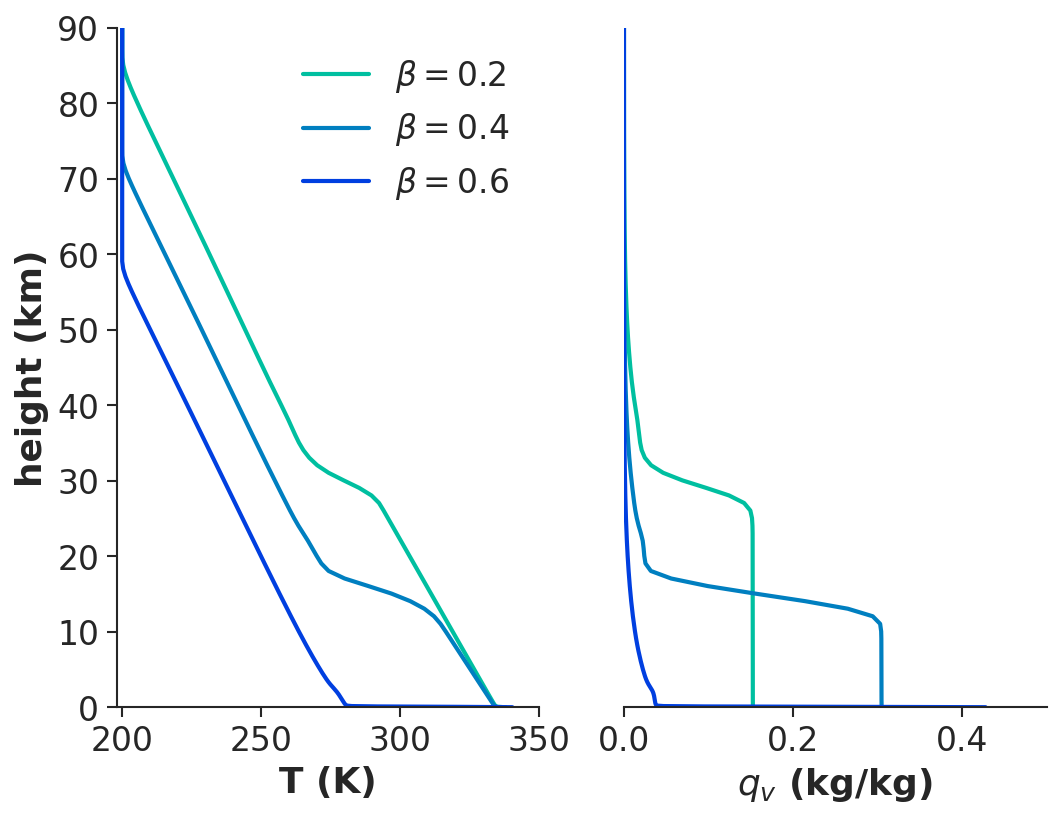}}
\caption{Results from the \textit{VaryBeta} experiment (Left) Temperature profiles. (Right) Specific humidity profiles.}
\label{fig:varyBeta}
\end{figure}

In the Guillot regime, how deep must the dry subcloud layer be to prevent the superadiabatic layer from merging into the surface? We investigated this question with the \textit{VaryBeta} experiment, which was branched from the \textit{DeepBL} case with $\mu_a=4$ g/mol but with surface wetness parameter $\beta$ increased from 0.2 to 0.4 and 0.6; this has the effect of increasing the near-surface humidity and thereby lowering the height of the lifting condensation level. For the case with $\beta=0.4$, the depth of the dry subcloud layer is approximately cut in half, to about 15 km, but the superadiabatic layer remains aloft (Fig. \ref{fig:varyBeta}, left panel). However, for $\beta=0.6$, the dry subcloud layer collapses and the superadiabatic layer merges with the surface, significantly reducing the total column water vapor (i.e, $\int \rho q_v\ \text{d}z$) from more than 800 kg/m$^2$ (for $\beta=0.4$) to less than 100 kg/m$^2$ (Fig. \ref{fig:varyBeta}, right panel). The \textit{VaryBeta} experiment thus points to a counter-intuitive conclusion: in the Guillot regime, increasing the surface moisture availability (e.g., by reducing the land fraction of the surface) can significantly \textit{decrease} overall atmospheric humidity.

\begin{figure}[ht!]
\centerline{\includegraphics[width=6in]{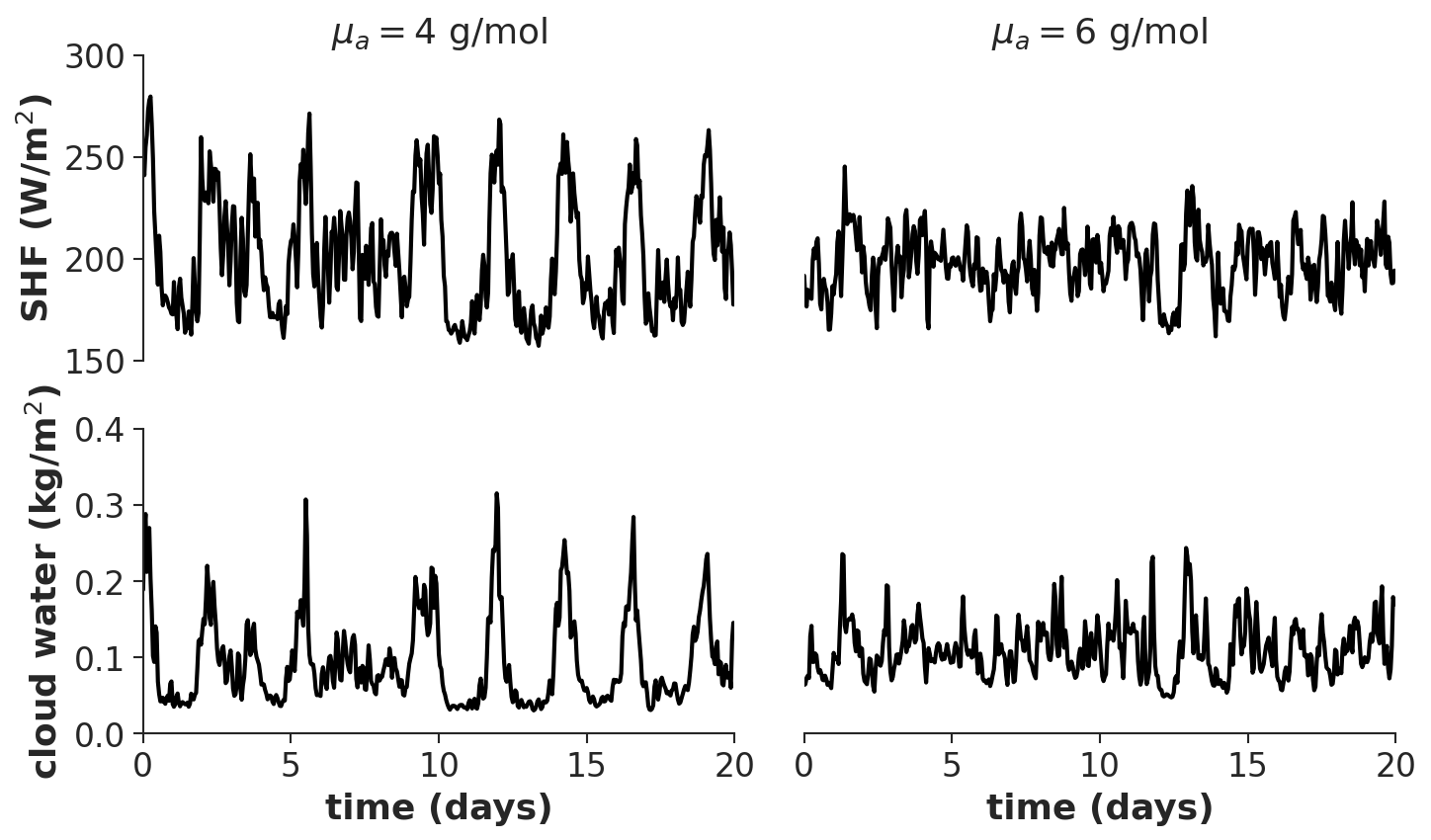}}
\caption{High-frequency (hourly) output from 20 days of the \textit{DeepBL} simulations with $\mu_a=4$ and 6 g/mol. Top row: surface sensible heat flux (SHF) in W/m$^2$. Bottom row: column-integrated cloud water in kg/m$^2$.}
\label{fig:episodic}
\end{figure}

An additional intriguing aspect of our simulations concerns the temporal variability of convection. Some of our simulations that are in the Guillot regime exhibit emergent periodicity in convective activity: for example, convection in the elevated moist layer of the \textit{DeepBL} case with $\mu_a=4$ g/mol ``pulses", with a period of about 2 days (Fig. \ref{fig:episodic}). However, this episodicity is not a defining feature of the Guillot regime; the \textit{DeepBL} case with $\mu_a=6$ g/mol does not exhibit oscillatory convective activity (Fig. \ref{fig:episodic}, right column). These results add to a growing set of experimental configurations known to produce episodic convection \citep{Seeley2021,Dagan2023,Song2024,Spaulding-Astudillo2024}, and further motivate the development of a general theory for when a radiative-convective equilibrium state is steady versus oscillatory.

\section{Discussion and conclusions}\label{sec:disc}
Our results add to a growing body of three-dimensional simulations of compositionally-inhibited convection. \cite{Clement2024} used cloud-resolving simulations to investigate convective inhibition due to methane condensation on Uranus and Neptune, and confirmed the existence of stably stratified superadiabatic layers when the methane abundance is above the Guillot threshold. Intriguingly, their simulations also revealed intermittent or episodic convective events similar to the behavior we document in our Figure \ref{fig:episodic}. The study by \cite{Ge2024} used a cloud-resolving model that can simulate multiple condensing species to explore the combined influence of CH$_4$ and H$_2$S on Neptune's atmosphere, and found that each condensible produces its own stably stratified superadiabatic layer. In addition, the recent study of \cite{Leconte2024} used a cloud-resolving model to simulate convective inhibition in the context of the temperate Neptune-like exoplanet K2-18b. The temperature profiles they obtain are qualitatively very similar to ours (e.g., their Figure 1), including the existence of the thin dry-adiabatic layer between the superadiabatic layer and the moist troposphere above. Overall, our idealized simulation framework that probes the parameter space between the Earthlike and hydrogen-dominated convective regimes is complementary to these other studies which are tailored to specific planetary contexts, bolstering confidence in the generality of the underlying physics.

One key takeaway from our parameter sweep of $\mu_a$ is that crossing into the regime with $q_s \ge q_\mathrm{crit}$ is associated with a stark increase in cloudiness, whether the atmosphere has an oceanic lower boundary condition (Fig. \ref{fig:guillot}) or suppressed surface evaporation and deep-atmosphere moisture availability (Fig. \ref{fig:deepBL}). Since younger planets are expected to have larger H$_2$ inventories \citep{Pierrehumbert2011,wordsworth2012transient,mol2022potential,Young2023}, this implies a potentially observable systematic trend toward increased cloudiness for younger exoplanets with low mean molecular weight, all else being equal. This result should be further validated with additional simulations using a fully-interactive radiative transfer scheme. Such simulations will be able to better capture the effect of changes in static stability (e.g., at the transition between the superadiabatic layer and the moist troposphere) on radiative cooling rates and clear-sky convergence, which are known to play a role in determining cloud fraction \citep[e.g.,][]{Bony2016a}.

Would the superadiabatic cloudy layers simulated in this work be observable on an exoplanet? In our \textit{DeepBL} simulations, the extensive cloud decks lie well below the $p=0.1$ bar level (Fig. \ref{fig:deepBL}), which we can take as a rough indication that such layers may not be observable in transit spectroscopy. Under what thermodynamic conditions would these cloud decks become accessible to remote observations? We can begin to address this question by solving for the conditions under which $q_v^*(p,T)=q_\mathrm{crit}$, with $p=0.1$ bar. This analysis reveals that, with H$_2$O as the condensible and with $\mu_a=2.5$ g/mol (i.e., assuming solar composition), the superadiabatic layer would occur at the $p=0.1$ bar level if that level had a temperature of about 247 K. This is quite close to present-day Earth's emission temperature (255 K), indicating that a superadiabatic cloudy layer could be a dominant influence on the transmission/emission spectra of planets with solar-composition background atmospheres and Earth-like emission temperatures.

\begin{figure}[ht!]
\centerline{\includegraphics[width=4.5in]{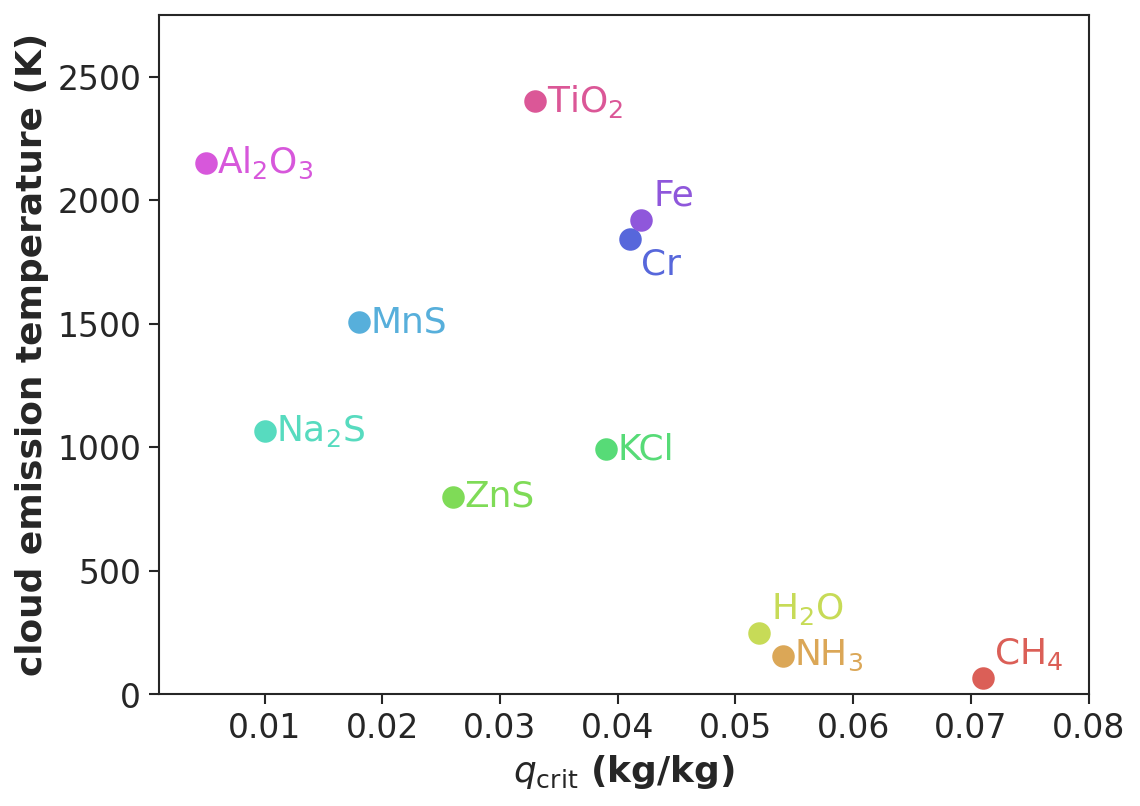}}
\caption{Cloud emission temperatures and $q_\mathrm{crit}$ values for a wide range of condensing species. We define cloud emission temperature as the temperature for which $q_v^*(p,T)=q_\mathrm{crit}$, for a representative emission pressure of $p=0.1$ bar. The latent heats and saturation vapor pressure curves required to calculate $q_\mathrm{crit}$ and $q_v^*$ were obtained from \cite{Gao2020} and \cite{Pierrehumbert2011_book}.}
\label{fig:condensibles}
\end{figure}

Although our simulations modeled water vapor as the condensible species, the convective physics of the Guillot regime is general and should also apply to more exotic condensibles, provided the species in question is sufficiently abundant in the lower atmosphere. As a rough guide to when superadiabatic cloudy layers might be accessible to remote observations, we can repeat the above analysis for a wide range of condensibles expected in planetary atmospheres (i.e., finding the temperature at which $q_v^*(p,T)=q_\mathrm{crit}$, for $p=0.1$ bar; we refer to this temperature as the ``cloud emission temperature'', i.e., the temperature at which the superadiabatic cloudy layer might be in the optically-thin part of the atmosphere as viewed from space.) The results are shown in Figure \ref{fig:condensibles} and summarized in Table \ref{tab:condensibles}. For a planet with a condensible species and brightness/emission temperature as indicated by Table \ref{tab:condensibles}, extensive superadiabatic cloud decks may be a dominant influence on their transmission/emission spectra if the condensible species is sufficiently abundant. Some of the condensibles we consider may be unlikely to exist at such high abundances: for example, \cite{kempton2011atmospheric} estimated the equilibrium abundance of \ce{KCl} to be roughly 1.5 parts per million in GJ 1214b's atmosphere (at 0.5 bar, 800 K, and assuming 30$\times$ solar metallicity), which is less than the $q_\mathrm{crit}$ value for \ce{KCl} in Figure \ref{fig:condensibles} and Table \ref{tab:condensibles}. A more thorough investigation of the feasibility of attaining condensible abundances that approach $q_\mathrm{crit}$ values under diverse planetary conditions is warranted.

An increased presence of clouds also has climate implications. If the albedos of exoplanets with thin \ce{H2}-dominated atmospheres are systematically higher, that might partially mitigate the intense warming effect from the collision-induced \ce{H2}-\ce{H2} greenhouse. However, as the \ce{H2} envelope is gradually lost to space, the associated increase in $\mu_a$ could result in a dramatic shock to climate when $q_s$ falls below $q_\mathrm{crit}$. The sudden disappearance of the superadiabatic cloud layers and restructuring of the tropospheric temperature profile would have large effects on the radiative balance of a planet, potentially causing very large changes in surface temperature on a short timescale compared to the planet's lifetime. Future work should investigate the radiative impacts of superadiabatic cloud layers in \ce{H2}-rich atmospheres across the model hierarchy, and ideally using a comprehensive radiative transfer model coupled to a global cloud-resolving model.

\begin{table}[]
\centering
\begin{tabular}{c|l|l|l}
\textbf{condensible} & \textbf{$\mu_v$ (g/mol)} & \textbf{$q_\mathrm{crit}$ (kg/kg)} & \textbf{cloud emission temperature (K)} \\
\hline
CH$_4$      & 16       &0.071&        65.2                          \\
NH$_3$      & 17       &0.054&       156.7                          \\
H$_2$O      & 18       &0.052&        247.3                          \\
ZnS         & 97.47    &0.026&        935.2                          \\
KCl         & 74.55    &0.039&        991.8                          \\
Na$_2$S     & 78.04    &0.01&        1065.5                         \\
MnS         & 87       &0.018&        1505.6                         \\
Cr          & 51.99    &0.041&        1841.8                         \\
Fe          & 55.84    &0.042&        1918.6                         \\
Al$_2$O$_3$ & 101.96   &0.005&        2031.4                         \\
TiO$_2$     & 79.86    &0.033&        2400.9                        
\end{tabular}
\caption{Molar masses and cloud emission temperatures for a wide range of condensing species. We define cloud emission temperature as the temperature for which $q_v^*(p,T)=q_\mathrm{crit}$, for a representative emission pressure of $p=0.1$ bar. The latent heats and saturation vapor pressure curves required to calculate $q_\mathrm{crit}$ and $q_v^*$ were obtained from \cite{Gao2020} and \cite{Pierrehumbert2011_book}.}
\label{tab:condensibles}
\end{table}

\begin{acknowledgments}
\textbf{Acknowledgments:} The authors acknowledge
funding from NSF award AGS2210757, and thank Diana Powell for help with the thermodynamics of exoplanet condensibles.

\textbf{Data availability:} The cloud-resolving model output and the code that generates the figures in this manuscript is available at \url{https://doi.org/10.5281/zenodo.14279238}.
\end{acknowledgments}

%

\vspace{5mm}





\appendix

\setcounter{table}{0}
\renewcommand{\thetable}{A\arabic{table}}
\renewcommand*{\theHtable}{\thetable}

\begin{table}[]
\centering
\begin{tabular}{ccc|cc|cc|cc|cc|cc}
 & \multicolumn{2}{c|}{\textit{\textbf{VaryMu}}} & \multicolumn{2}{c|}{\textit{\textbf{VaryMu\_280}}} & \multicolumn{2}{c|}{\textit{\textbf{VaryMu\_320}}} & \multicolumn{2}{c|}{\textit{\textbf{DeepBL}}} & \multicolumn{2}{c|}{\textit{\textbf{VaryBeta}}} & \multicolumn{2}{c}{\textit{\textbf{DeepBL\_hr}}} \\
\multicolumn{1}{c|}{$\mu$ (g/mol)} & $L_z$ (km) & $N_z$ & $L_z$ (km) & $N_z$ & $L_z$ (km) & $N_z$ & $L_z$ (km) & $N_z$ & $L_z$ (km) & $N_z$ & $L_z$ (km) & $N_z$ \\ \hline
\multicolumn{1}{c|}{28} & 24.07  & 55  & 19.07  & 50  & 31.07  & 62  & -      & -   & -     & - & - & - \\
\multicolumn{1}{c|}{26} & 25.07  & 56  & 20.07  & 51  & 37.07  & 68  & -      & -   & -     & - & - & - \\
\multicolumn{1}{c|}{24} & 26.07  & 57  & 21.07  & 52  & 39.07  & 70  & -      & -   & -     & - & - & - \\
\multicolumn{1}{c|}{22} & 27.07  & 58  & 22.07  & 53  & 42.07  & 73  & -      & -   & -     & - & - & - \\
\multicolumn{1}{c|}{20} & 29.07  & 60  & 23.07  & 54  & 45.07  & 76  & -      & -   & -     & - & - & - \\
\multicolumn{1}{c|}{18} & 31.07  & 62  & 24.07  & 55  & 48.07  & 79  & -      & -   & -     & - & - & - \\
\multicolumn{1}{c|}{16} & 34.07  & 65  & 26.07  & 57  & 53.07  & 84  & -      & -   & -     & - & - & - \\
\multicolumn{1}{c|}{14} & 37.07  & 68  & 28.07  & 59  & 58.07  & 89  & 63.07  & 94  & -     & - & - & - \\
\multicolumn{1}{c|}{12} & 42.07  & 73  & 31.07  & 62  & 62.07  & 93  & 70.07  & 101 & -     & - & - & - \\
\multicolumn{1}{c|}{10} & 48.07  & 79  & 35.07  & 66  & 66.07  & 97  & 79.07  & 110 & -     & - & - & - \\
\multicolumn{1}{c|}{9}  & 52.07  & 83  & -      & -   & -      & -   & -      & -   & -     & - & - & - \\
\multicolumn{1}{c|}{8}  & 57.07  & 88  & 42.07  & 73  & 79.07  & 110 & 92.07  & 123 & -     & - & - & - \\
\multicolumn{1}{c|}{7}  & 64.07  & 95  & -      & -   & -      & -   & -      & -   & -     & - & - & - \\
\multicolumn{1}{c|}{6}  & 72.07  & 103 & 52.07  & 83  & 99.07  & 130 & 98.07  & 129 & -     & - & 96.75 & 250 \\
\multicolumn{1}{c|}{5}  & 84.07  & 115 & -      & -   & -      & -   & -      & -   & -     & - & - & - \\
\multicolumn{1}{c|}{4}  & 102.07 & 133 & 73.07  & 104 & 137.07 & 168 & 134.07 & 165 & 99.07 & 130 & - & - \\
\multicolumn{1}{c|}{3}  & 131.07 & 162 & -      & -   & -      & -   & -      & -   & -     & - & - & - \\
\multicolumn{1}{c|}{2}  & 187.07 & 218 & 135.07 & 166 & 237.07 & 268 & -      & -   & -     & - & - & - 
\end{tabular}
\caption{Vertical dimension ($L_z$) and number of vertical levels ($N_z$) in the simulations. For compactness of presentation, $L_z$ is rounded to the nearest 10 meters.}
\label{tab:zdims}
\end{table}


\bibliography{main}{}
\bibliographystyle{aasjournal}



\end{document}